\documentclass[reprint,amssymb, amsmath, aps, superscriptaddress, showpacs, footinbib, prl]{revtex4-1}
\usepackage[T1]{fontenc}
\usepackage[utf8]{inputenc}
\usepackage{amssymb}
\usepackage{amsmath}
\usepackage{color}
\usepackage{graphicx}
\usepackage{footmisc,booktabs}
\usepackage{gensymb}
\usepackage{sidecap}
\usepackage{microtype}

\begin{document}

\title{NaYbS$_2$\,--\,a new planar spin-1/2 triangular-lattice magnet and putative spin liquid}

\author{M.~Baenitz}\email[Corresponding author: \vspace{5pt}]{Michael.Baenitz@cpfs.mpg.de}
\affiliation{Max Planck Institute for Chemical Physics of Solids, D-01187 Dresden, Germany}

\author{Ph.~Schlender}
\affiliation{ Faculty of Chemistry and Food Chemistry, TU Dresden, D-01062 Dresden, Germany}

\author{J.~Sichelschmidt}
\affiliation{Max Planck Institute for Chemical Physics of Solids, D-01187 Dresden, Germany}

\author{Y.~A. Onykiienko}
\affiliation{Institute  for Solid State and Materials Physics, TU Dresden, D-01062 Dresden, Germany}

\author{Z.~Zangeneh}
\affiliation{Leibniz Institute for Solid State and Materials Research, IFW Dresden,  D-01069 Dresden, Germany}

\author{K.~M. Ranjith}
\affiliation{Max Planck Institute for Chemical Physics of Solids, D-01187 Dresden, Germany}

\author{R. Sarkar}
\affiliation{Institute  for Solid State and Materials Physics, TU Dresden, D-01062 Dresden, Germany}

\author{L.~Hozoi}
\affiliation{Leibniz Institute for Solid State and Materials Research, IFW Dresden, D-01069 Dresden, Germany}

\author{H.~C. Walker}
\affiliation{ISIS Neutron and Muon Source, Rutherford Appelton Laboratory, Chilton, Didcot OX11 OQX, United Kingdom}

\author{J.-C. Orain}
\affiliation{Laboratory for Muon Spin Spectroscopy, Paul Scherrer Institute, CH-5232 Villingen PSI, Switzerland  }

\author{H.~Yasuoka}
\affiliation{Max Planck Institute for Chemical Physics of Solids, D-01187 Dresden, Germany}

\author{J. van den Brink}
\affiliation{Leibniz Institute for Solid State and Materials Research, IFW Dresden, D-01069 Dresden, Germany}


\author{H.~H. Klauss}
\affiliation{Institute  for Solid State and Materials Physics, TU Dresden, D-01062 Dresden, Germany}

\author{D.~S. Inosov}
\affiliation{Institute for Solid State and Materials Physics, TU Dresden, D-01062 Dresden, Germany}

\author{Th.~Doert}
\affiliation{ Faculty of Chemistry and Food Chemistry, TU Dresden, D-01062 Dresden, Germany}

\date{\today}

\begin{abstract}\noindent
Platelike high-quality NaYbS$_{2}$ rhombohedral single crystals with lateral dimensions of a few mm have been grown and investigated in great detail by bulk methods like magnetization and specific heat, but also by local probes like nuclear magnetic resonance (NMR), electron-spin resonance (ESR), muon-spin relaxation ($\mu$SR), and inelastic neutron scattering (INS) over a wide field and temperature range. Our single-crystal studies clearly evidence a strongly anisotropic quasi-2D magnetism and an emerging spin-orbit entangled $S=1/2$ state of Yb towards low temperatures together with an absence of long-range magnetic order down to 260~mK. In particular, the clear and narrow Yb ESR lines together with narrow $^{23}$Na NMR lines evidence an absence of inherent structural distortions in the system, which is in strong contrast to the related spin-liquid candidate YbMgGaO$_{4}$ falling within the same space group $R\overline{3}m$. This identifies NaYbS$_{2}$ as a rather pure spin-1/2 triangular lattice magnet and a new putative quantum spin liquid.
\end{abstract}


\maketitle

\textit{Introduction}.\,---\,In low-dimensional quantum magnets, competing confined magnetic exchange interactions restrict the magnetic degrees of freedom, which leads to a strong frustration accompanied by enhanced quantum fluctuations. Ultimately this prevents the systems from long-range order, and the ground state is supposed to be a magnetic liquid. There are various types of such quantum spin liquids (QSL) depending on the lattice geometry (in 2D: square-, triangular-, kagome-, or honeycomb-type; in 3D: hyperkagome, hyperhoneycomb, or pyrochlore), the magnetic exchange (e.g. Heisenberg, Kitaev, or Dzyaloshinskii-Moriya type), and the magnetic ion itself \cite{Lee, Balents, Savay, Zhou}. Planar spin-$1/2$ triangular lattice magnets (TLMs) with antiferromagnetic exchange interactions are ideal QSL candidates as proposed by P.~W. Anderson \cite{Anderson}. A few examples are found among the organic materials, such as K-(BEDT-TTF)$_{2}$Cu$_{2}$(CN)$_{3}$ \cite{Shimizu} and EtnMe$_{4-n}$Sb[Pd(DMIT)$_{2}$]$_{2}$ \cite{Yamashita1}, whereas among inorganic compounds such QSL model systems are very rare, e.g. Ba$_{3}$CuSb$_{2}$O$_{9}$ \cite{Zhou1}.

For TLMs the fingerprint of the emerging U(1)-QSL state is the formation of the gapless spinon Fermi surface evidenced by a finite and constant magnetic specific heat coefficient $C_{\rm m}/T$ and a finite residual magnetic susceptibility \cite{Balents, Savay, Zhou, Sik}. Persisting low-energy magnetic excitations with characteristic dispersion relations could be evidenced by inelastic neutron scattering (INS), nuclear magnetic resonance (NMR) and muon spin relaxation ($\mu$SR) \cite{Balents, Savay, Zhou, Shimizu, Itou, Shen}. Recently, the field of $S=1/2$ quantum magnetism was extended away from 3$d$ ions (like Cu$^{2+}$ and V$^{4+}$) towards 4$d$, 5$d$, and even 4$f$ systems~\cite{Balents}. In these materials, an effective $J_{\rm eff}=1/2$ moment can be realized due to strong spin-orbit coupling (SOC). The honeycomb 4$d$ and 5$d$ QSL candidates $\alpha$-RuCl$_{3}$ and Na$_{2}$IrO$_{3}$ are $proximate \ QSLs$ \cite{Trebst} and exhibit long-range order, whereas the 4$f$-TLM YbMgGaO$_{4}$ is claimed to be a gapless QSL with an absence of order and persistent magnetic excitations down to the lowest temperatures \cite{Li, Li1, Li2, Shen, Paddison, Li3}. The strong spin-orbit interaction could be at the same energy scale as the Coulomb interaction $U$ and the crystal electric field (CEF) splitting, which finally leads to highly degenerate band-like magnetic states with complex excitations \cite{Knolle}. Especially for the 4$f$-TLMs (based on Ce or Yb), the spin-orbit entanglement leads to highly anisotropic interactions among the moments, which should strongly enhance quantum fluctuations and promote a QSL ground state \cite{Hu, Iqbal, Li4, Gong, Iaconis, Rau}. In the absence of SOC, the classical Heisenberg model predicts the energy-minimum solution to be the planar 120$^\circ$ N\'eel-ordered state with a strong magnetic anisotropy \cite{Kawamura, Chubukov, Yamamoto}. As the first QSL-TLM with a strong 4$f$-driven SOC, YbMgGaO$_{4}$ gained a lot of attention, but it turned out that there is a considerable site mixing between Ga and Mg ions, which affects the magnetic properties in general and the predicted QSL ground state in particular \cite{Androja, Kimchi}. As for Yb-TLMs, the SOC of the Yb ion and the CEF create a ground state doublet which at low temperatures could be described with an effective spin of $1/2$, any structural distortion consequently alters the CEF splitting and may strongly affect the magnetic ground state~\cite{Kimchi, Li3}.

\textcolor{black}{In the search for the ideal spin-$1/2$ TLM with spin-orbit interaction, we focused on the $A^{1+}R^{3+}X_{2}$ delafossites, where $A$ is a nonmagnetic monovalent metal ion, $R$ is a rare-earth ion (Ce or Yb), and $X$ stands for either oxygen or sulfur.} Most of them form in the same $R\overline{3}m$ space group as YbMgGaO$_{4}$, and most importantly they exhibit perfect triangular layers of $R^{3+}$ ions, composed of edge-sharing $R$O$_{6}$ octahedra. In contrast to YbMgGaO$_{4}$, the $R^{3+}$ ions in the delafossite structure are at the inversion center of the lattice. Furthermore, delafossites have an ABAB stacking of the triangular layers along the $c$ axis, in contrast to the \mbox{ABCABC} stacking in YbMgGaO$_{4}$. Among the delafossites, there are some early reports on $A$YbO$_{2}$ ($A$~=~Ag,\,Na) that suggest a pseudo-spin $S=1/2$ ground state \cite{Hashimoto, Saito, Miyasaka}. So far no single crystals were accessible for these systems, whereas there is a report about the single crystal growth of the sulfur homologue NaYbS$_{2}$~\cite{Schleid}. We have grown sizable platelike high-quality single crystals and investigated them with bulk methods, such as magnetization and specific heat, but also with local probes like NMR, electron spin resonance (ESR) and $\mu$SR in great detail over a wide field and temperature range. Our study of NaYbS$_{2}$ single crystals, presented here, evidences strongly anisotropic quasi-2D magnetism and an emerging spin-orbit entangled $S=1/2$ state of Yb towards low temperatures together with an absence of long-range magnetic order down to 260~mK. In particular the clear and narrow Yb ESR lines together with narrow $^{23}$Na NMR lines evidence the absence of inherent structural distortions in the system, which is in strong contrast to YbMgGaO$_{4}$ \cite{Androja} and therefore identifies NaYbS$_{2}$ as a rather pure TLM and a new putative quantum spin liquid.

\textit{Experimental~techniques}.\,---\,We have grown NaYbS$_{2}$ single crystals by a modified method following Lissner and Schleid \cite{Schleid}, starting from rare-earth metal grains, sulfur, and sodium chloride as flux. After the reactions the water-insoluble product was washed with water and ethanol and dried at 60\,$\degree$C. The single crystals form as transparent yellowish thin ($\sim$\,100~$\mu$m) platelets with lateral dimensions up to 10~mm. For polycrystalline samples of NaYbS$_{2}$ and NaLuS$_{2}$, a salt metathesis was used, starting from rare-earth trichloride and sodium sulfide ground together with excess of sodium chloride as the flux, as detailed in the Supplemental Material (SM) \cite{SM}. NaYbS$_{2}$ forms in the rhombohedral  $\alpha$-NaFeO$_{2}$ delafossite structure ($R\overline{3}m$) with $a=3.895(1)$\,\AA\ and $c=19.831(6)$\,\AA\ \cite{Schleid, Range, Cotter}. Magnetization measurements were performed with SQUID magnetometers (MPMS, VSM) and an ac/dc susceptometer (PPMS) from Quantum Design. The MPMS was equipped with a $^3$He cooling stage (down to 500~mK) and the VSM was equipped with a goniometer to probe the angular dependence of the magnetization \cite{SM}. Specific heat measurements were conducted with a commercial PPMS from Quantum Design down to 350~mK. NMR measurements were carried out by applying conventional pulsed NMR in the field-sweep mode on both powder and single crystals \cite{SM}. $\mu$SR experiments down to 260~mK were performed at the Paul Scherrer Institute (DOLLY instrument) on sandwiched NaYbS$_{2}$ single crystals (48~mg). INS measurements were performed on the polycrystalline NaYbS$_{2}$ material (6~g) at the thermal-neutron time-of-flight (TOF) spectrometer MERLIN at ISIS Neutron and Muon Source. ESR experiments were performed at X- and Q-band frequencies (9.4 and 34~GHz) on single crystals and powders down to liquid helium temperatures \cite{Joerg}.

\textit{Results}.\,---\,Figure 1(a) shows the susceptibility of NaYbS$_{2}$ as a function of temperature in 0.1 and 7~T for the fields $H$ applied in the $(a,b)$ plane, $\chi_\perp(T,H)$, and along the $c$ direction, $\chi_\parallel(T,H)$. Above 80~K the magnetic anisotropy disappears, and the field dependence becomes negligible. Here the susceptibility $\chi(T)$ could be fitted well (after the subtraction of a small $T$-independent diamagnetic contribution $\chi_0$) with a Curie-Weiss (CW) law, which yields a Weiss temperature of $\theta = -65$~K and an effective moment of $\mu_{\rm eff} = 4.5\mu_{\rm B}$ [Fig.~1(b)]. The CW fit parameters are similar to those obtained from polycrystalline powder \cite{SM}, and the $\mu_{\rm eff}$ value agrees with the theoretical prediction for trivalent Yb with $J=7/2$ ($4.54\mu_{\rm B}$). Furthermore these values are rather similar to findings on AgYbO$_{2}$ powder samples \cite{Miyasaka}.
\begin{figure}[t]
\includegraphics[clip,width=\columnwidth]{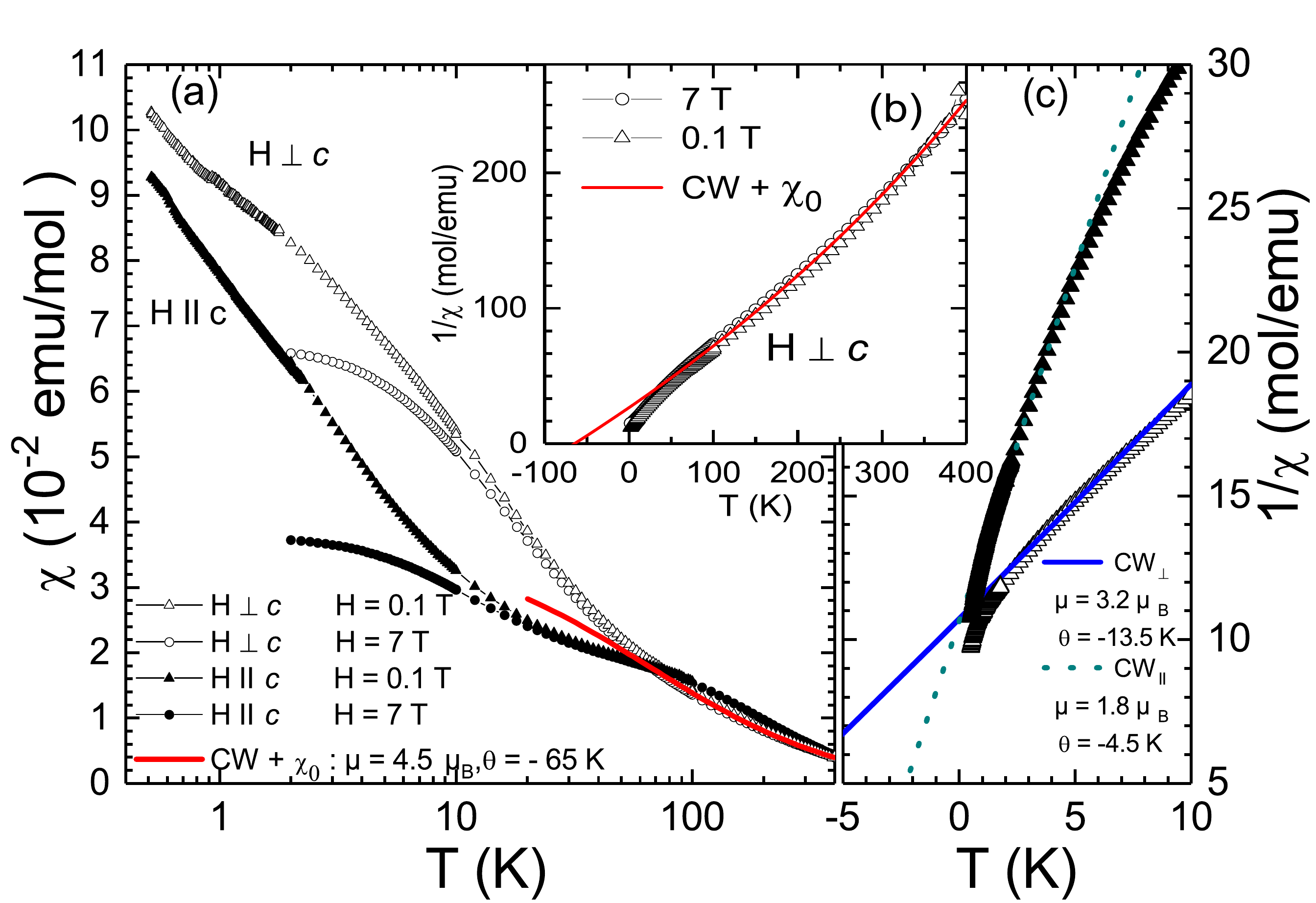}
\caption{(a)~Temperature dependence of the magnetic susceptibility ($\chi_\parallel$ and $\chi_\perp$) of NaYbS$_2$ measured in $H=7$~T and 0.1~T. The solid line corresponds to an extended Curie-Weiss fit which includes an additional small diamagnetic offset contribution ($\chi_0 = -0.0015$~emu/mol). (b)~Inverse susceptibility 1/$\chi_\perp$ and extended CW fit to the data between 80 and 400~K (solid line). (c) Inverse susceptibilities  below 10~K and CW fits to the data (solid and dotted lines).}
\end{figure}
Below 80~K a sizable magnetic anisotropy develops, and there is a strong field effect on the susceptibility which indicates an emerging effective low-spin state with a small exchange coupling.  Plotting the inverse susceptibility $1/\chi_\perp(T)$ (measured down to 0.5~K) at 0.1~T below 10~K clearly shows a CW behavior with $\mu_{\rm eff\perp}$= $3.2\mu_{\rm B}$ and  $\theta_\perp = -13.5$~K [Fig~1(c)] down to about 1~K. Below 1 K, 1/$\chi_\perp(T)$ bends over into a behavior with even smaller moment. For the field in the $c$ direction, below 10~K $1/\chi_\parallel(T)$ is still curved, but fitting a CW law below approximately 5~K provides a moment of $\mu_{\rm eff\parallel}=1.8\mu_{\rm B}$, which is close to the $S=1/2$ spin-only moment of $1.72\mu_{\rm B}$, and a Weiss temperature of $\theta_\parallel = -4.5$~K. The low-$T$ susceptibility clearly evidences a SOC entangled enhancement of the magnetic moment in plane ($3.2\mu_{\rm B}$), whereas out of plane the CW moment is close to the spin only value. Below 1~K it seems that both $1/\chi(T)$ curves tend to merge. The fact that there is no magnetic order above 0.26~K allows an estimate of a lower limit for the frustration parameter which is $f = \theta_\perp$/(0.26~K)~=~52 in the ($a,b$) plane and $f=\theta_\parallel$/(0.26~K)~=~17 in the $c$-direction.

\begin{figure}[t]
\includegraphics[clip,width=0.8\columnwidth]{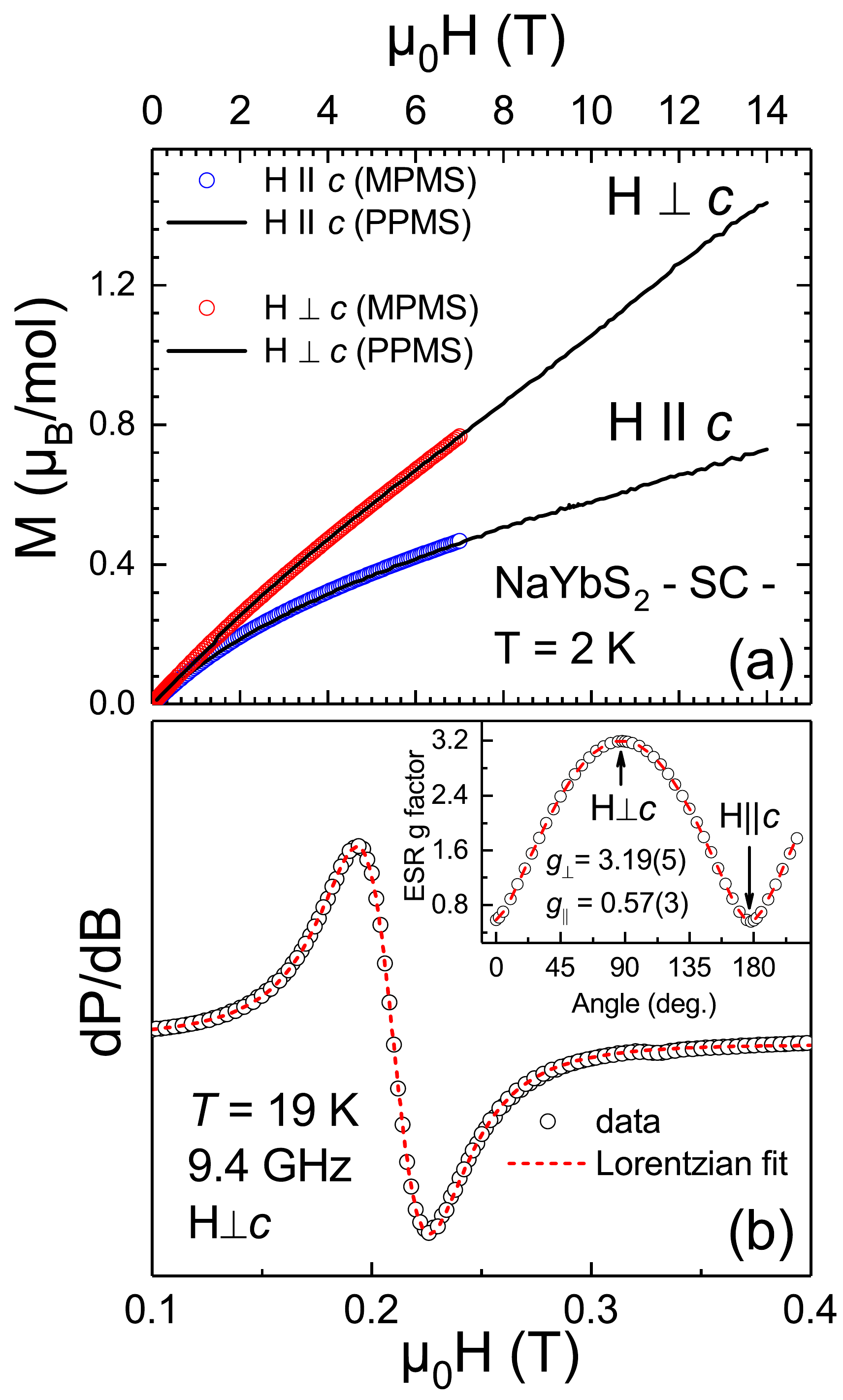}
\caption{(a) Magnetization versus field for $\mathbf{H}\parallel\mathbf{c}$ and $\mathbf{H}\perp\mathbf{c}$ at $T=2$~K (open symbols and solid lines represent data from SQUID measurements (Quantum Design MPMS) and from dc-susceptometer measurements (Quantum Design PPMS), respectively. (b) ESR line at 19 K (the dashed line corresponds to a Lorentzian fit). The inset shows the angular dependence of the ESR $g$-factor at 19~K (dashed line represents a uniaxial fit).}
\end{figure}

The magnetization $M(H)$ was measured for $\mathbf{H}\parallel \mathbf{c}$ and $\mathbf{H}\perp \mathbf{c}$ in fields up to 14~T at $T=2$~K (for powder data at 0.5~K see SM \cite{SM}). Both magnetizations show no sign of saturation in that field range (Fig.~2a). From the difference in the Weiss temperature and the anisotropy of $\mu_{\rm eff}$, the strongest field effect on $M(H)$ is expected for \mbox{$\mathbf{H}\parallel\mathbf{c}$}. As a further probe for the anisotropy we applied ESR on the NaYbS$_2$ single crystal. A well resolved and narrow Yb ESR line could be found for both orientations (Fig.~2b). This is in strong contrast to YbMgGaO$_{4}$, where structural distortions (Mg-Ga site mixing, Yb sits on noncentrosymmetric positions) lead to a CEF randomness resulting in a rather broad and much less resolved ESR line (with $g_\perp = 3.06$ and $g_\parallel=3.72$) and a strong broadening of inelastic-neutron CEF peaks \cite{Androja}. The ESR $g$-factor in NaYbS$_2$ is strongly anisotropic (with $g_\perp = 3.19$ and $g_\parallel = 0.57$), which signals a large CEF anisotropy at the Yb ion (Fig.~2b, inset). In general the $g$ factors describe the Zeeman splitting of the lowest Kramers doublet of the Yb ion and depend on the local site symmetry and the character of the ground-state wave function \cite{Abragam}. From the ESR $g$ values, saturation magnetizations of about $M_{\rm sat\perp} = 1.6\mu_{\rm B}$ and $M_{\rm sat\parallel}=0.285\mu_{\rm B}$ are expected. As seen in Fig.~2 (a), $M_{\rm sat\parallel}(H)$ exceeds this estimate by more than a factor of 2, whereas in the $(a,b)$ plane $1.6\mu_{\rm B}$ is not reached up to 14~T. For $\mathbf{H}\parallel\mathbf{c}$, $M(H)$ might be dominated by a Van Vleck contribution in the susceptibility, resulting in a linear behavior ($M_{\parallel c}$ $\approx$ $H \chi_{vv}$) towards high magnetic fields.

\begin{figure}[t]
\includegraphics[clip,width=\columnwidth]{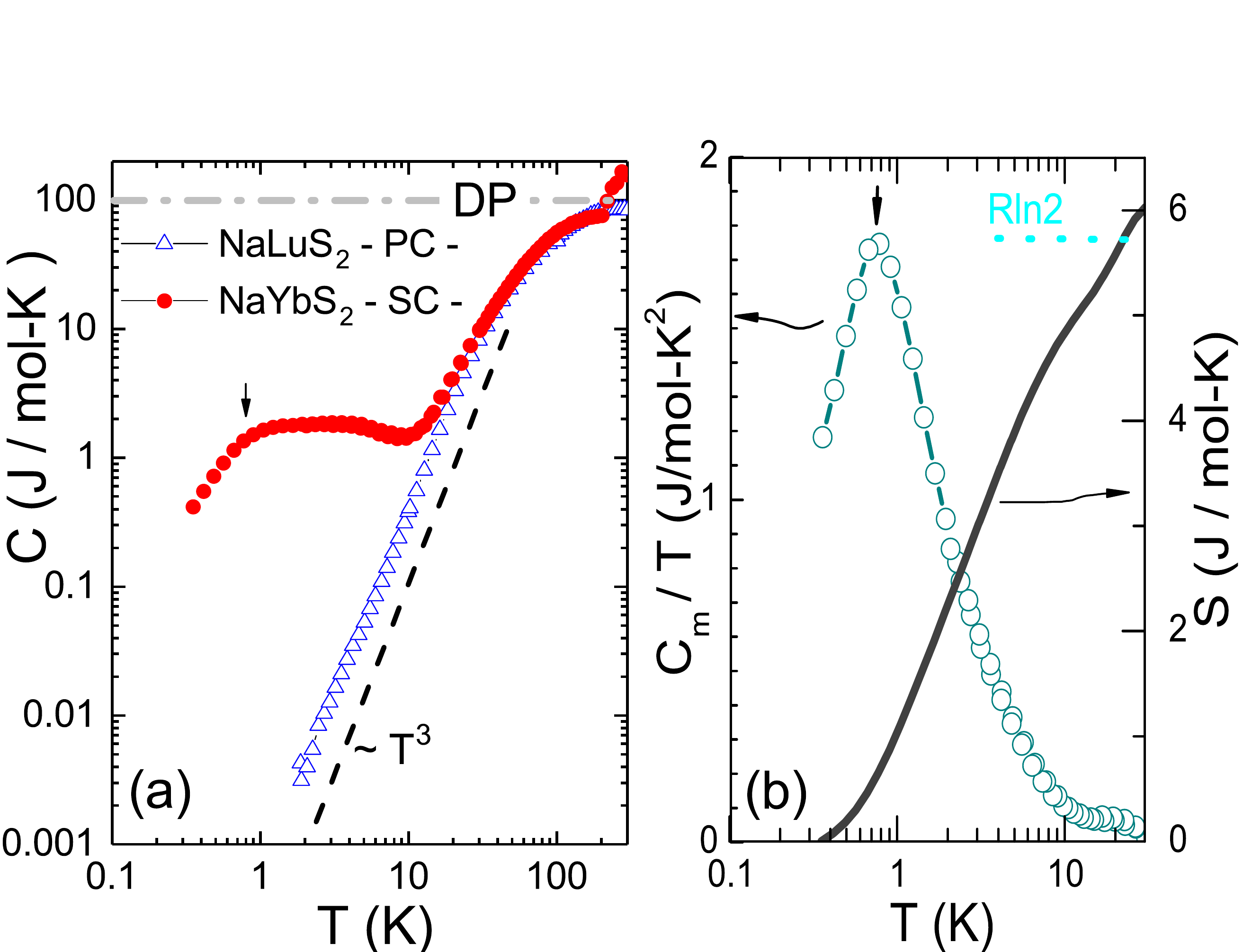}
\caption{(a) Temperature dependence of the specific heat of NaYbS$_2$ measured on stacked single crystals (SC) together with the non magnetic structural homologue NaLuS$_2$ (polycrystalline material (PC) as a pressed pellet). The horizontal dashed line indicates the Dulong Petit value (DP). (b) Magnetic heat capacity of NaYbS$_2$ divided by $T$ as a function of temperature (left axis) and calculated entropy (right axis).}
\end{figure}

Specific-heat measurements have been performed on two single-crystalline NaYbS$_2$ samples (stacked single crystals of 15~mg total mass (Fig.~3) and a single crystal with a mass of 0.2~mg~\cite{SM}) and polycrystalline NaLuS$_2$ as a nonmagnetic phonon reference between 350~mK and 300~K (Fig.~3a). The magnetic contribution to the specific heat $C_{\rm m}$ is obtained by the subtraction of the phonon reference from the data. Figure~3 (b) shows $C_{\rm m}/T$ together with the magnetic entropy. Although $C_{\rm m}/T$ vs. $T$ exhibits a peak at about $T^* = 0.8$~K, the entropy shows no anomaly at $T^*$ and increases smoothly towards higher $T$, merging with the spin-1/2 value of $S\approx R\ln2$ at about 20~K. The $T^*$-peak is rather narrow in comparison to the hump found in YbMgGaO$_{4}$ at 2~K that originates from a (disorder-induced) Schottky effect~\cite{Li}. As there are no signs of order in susceptibility (see Fig.~1) and $\mu$SR (see SM for results on stacked single crystals \cite{SM}), we speculate that the $T^*$ peak signals the emerging QSL state probably with a partial gapping out of magnetic excitations (see e.g. results on K-(BEDT-TTF)$_{2}$Cu$_{2}$(CN)$_{3}$ \cite{Yamashita}). Another scenario is that persistent spin fluctuations suppress magnetic order not completely, and the observed peak corresponds to a partial (probably short-range) magnetic order of a minor fraction of spins, whereas the majority part remains fluctuating. Nonetheless, in view of the susceptibility and preliminary $\mu$SR results, this appears unlikely. From the theory point of view the interaction between spin-orbit entangled Kramers-doublet local moments on planar triangles could be rather complex. Beside the classical 120$^\circ$ N\'eel phase and the QSL phase, complex magnetic textures like stripes (reminiscent of $\alpha$-RuCl$_{3}$) are predicted in the global phase diagram \cite{Li4, Gong}.

\begin{figure}[t]
\includegraphics[clip,width=\columnwidth]{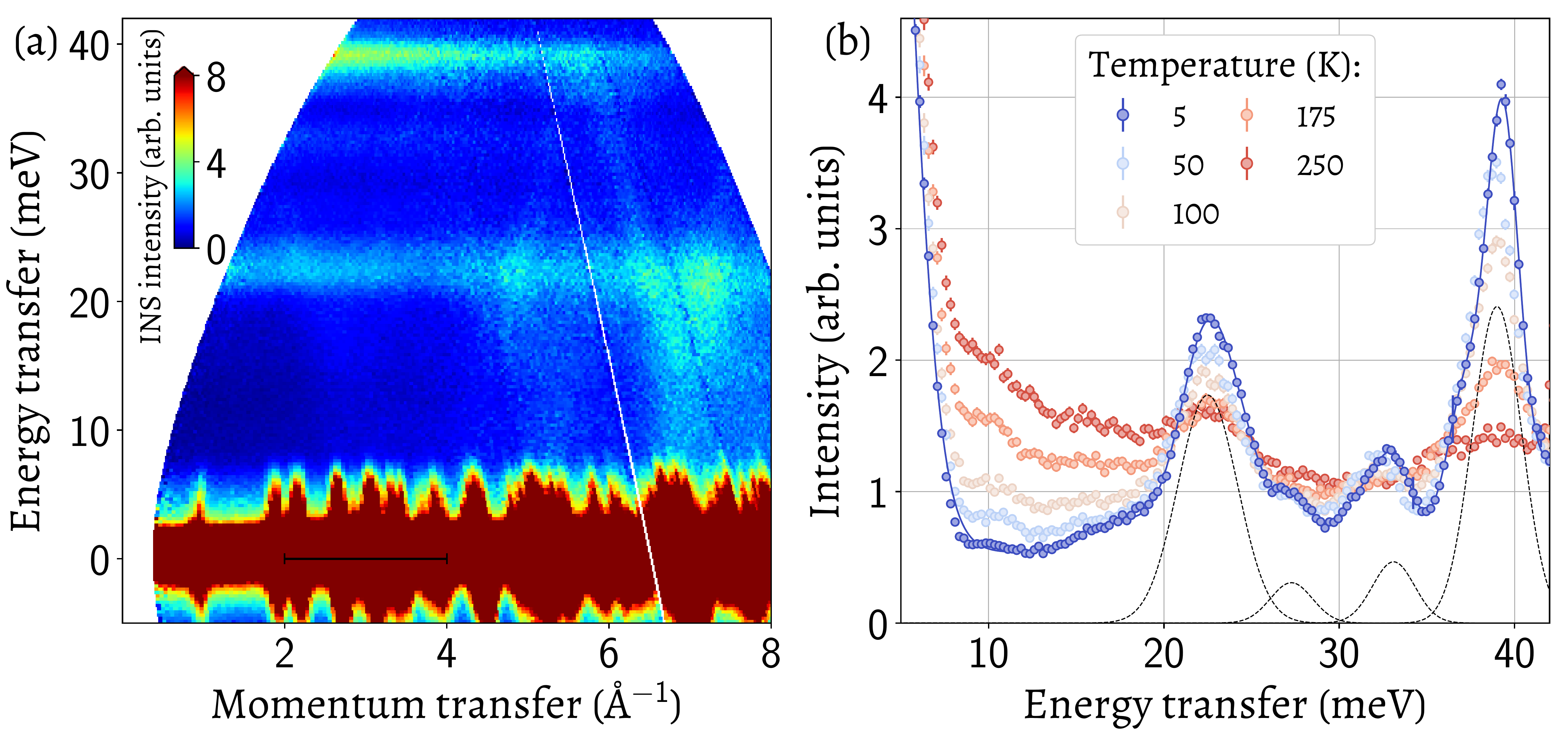}
\caption{(a) Powder TOF spectrum measured at 5~K with $E_ {\rm i} = 50$~meV. (b) The spectrum integrated over a momentum range of 2--4~\AA$^{-1}$ for various temperatures. The integration range is shown in (a) by the black interval. The black dashed line in (b) shows the individual peaks obtained from the fit of the 5~K data.}
\end{figure}

Further we probed the CEF excitations of the Yb$^{3+}$ ions by neutron spectroscopy. Figure~4(a) shows the TOF spectrum of the polycrystalline NaYbS$_2$ sample (6~g) at 5~K, measured with an incident neutron energy of $E_ {\rm i} = 50$~meV. Dispersing features that increase in intensity with increasing $|\mathbf{Q}|$ are associated to phonons, whereas horizontal lines with intensities decaying towards higher $|\mathbf{Q}|$ originate from CEF excitations. Figure~4(b) shows the integrated intensity between 2 and 4~\AA$^{-1}$ as a function of energy at several temperatures. The two most intense peaks at 23 and 39~meV share the same $T$-dependence, which suggests that they are related to the CEF excitations of the NaYbS$_2$ material. Additional less intense peaks at 27 and 32~meV might originate from a minority phase in the sample (see discussion in the SM \cite{SM}). Nonetheless in comparison with YbMgGaO$_{4}$ the INS peaks are rather narrow, which evidences the absence of inherent crystal-field randomness in agreement with the narrow ESR lines. Furthermore the CEF levels are found at lower energies compared to YbMgGaO$_{4}$~\cite{Li3}. Making use of the $E_ {\rm i} = 131$~meV channel, we observed the spectra up to 100~meV, but we did not find any additional peaks. From the $T$- dependent ESR line width, the first excited CEF level is expected to be around 17~meV \cite{Joerg}. To cross-check the values extracted from the ESR and INS measurements for the splittings among the low-lying Yb$^{3+}$ $f^{13}$ levels, we further performed {\it ab initio} quantum chemistry calculations using experimentally determined atomic positions (see details in SM \cite{SM}). We found low-lying excited states at 14, 20, and 47~meV with respect to the ground-state doublet, as well as computed ground-state $g$ factors of $g_\perp=3.66$ and $g_\parallel=0.60$. All these computational results are in reasonable agreement with the experimental INS and ESR data and provide a solid starting point for a more detailed analysis of the electronic structure of NaYbS$_2$.


\textit{Discussion}.\,---\,For spin-orbit coupled quasi-2D TLMs the extended XXZ model is established to capture the impact of the SOC \cite{Li2, Li5}. Here the interaction between the Kramers-doublet $S = 1/2$ moments is anisotropic in spin space and in real space. Therefore the interaction depends on the bond direction, which is a common feature among spin-orbit entangled QSLs (e.g. $\alpha$-RuCl$_{3}$ and Na$_{2}$IrO$_{3}$). A complete estimate of the exchange constants is not possible with the data set presented. Nonetheless, from the Weiss constants $\theta_\perp = 3J^{+-} = -13.5$~K and $\theta_\parallel = (3/2)J^{zz} = -4.5$~K one could roughly estimate the standard XXZ model out-of-plane exchange $J^{zz} = -3$~K and the in-plane exchange $J^{+-} = -4.5$~K. These values are both larger than those found for YbMgGaO$_{4}$ ($J^{zz}$ = $J^{+-} = -1$\,K) \cite{Li1}. The comparison of NaYbS$_2$ with YbMgGaO$_{4}$ is rather complex for the following reasons. First, the YbS$_{6}$ octahedron is larger than the corresponding YbO$_{6}$ octahedron in YbMgGaO$_{4}$ due to the difference in ionic radii of S$^{2-}$ and O$^{2-}$ and the absence of (Mg/Ga)O$_5$ bipyramids \cite{SM}. Second, the Yb layer distance along the $c$ axis is reduced for NaYbS$_2$ (6.57~\AA\ vs. 8.38~\AA\ for YbMgGaO$_{4}$), and the $a$ axis of NaYbS$_2$ is slightly larger which finally leads to a 3.5\% increase of the cell volume for NaYbS$_2$. Finally, in NaYbS$_2$ the Yb ion resides in the inversion center of the structure which has an impact on the magnetic exchange. From that we conclude that the rhombohedral lattice distortion in the trigonal system is much more prominent in NaYbS$_2$ than in YbMgGaO$_{4}$ and causes both, the $g$-factor anisotropy and the exchange anisotropy in the susceptibility. Furthermore, the $c/a$ ratio of 5.1 for NaYbS$_2$ is smaller than that of 7.4 in YbMgGaO$_{4}$, which in first view explains the relative shift of the CEF levels towards lower energies. In conclusion, our studies on NaYbS$_2$ clearly evidence a spin-orbit entangled anisotropic magnetism associated with the Kramers-doublet $S=1/2$ local moment and the absence of inherent structural distortions. The combination of single-crystal magnetization and ESR evidences an exchange anisotropy and a strong $g$-factor anisotropy. Furthermore single-crystal susceptibility, specific-heat, and $\mu$SR measurements evidence the absence of long-range magnetic order down to 260~mK. The planar TLM NaYbS$_2$ therefore could be regarded as a putative QSL candidate, and the data presented here will certainly stimulate more detailed future investigations in the mK regime on the complex ground state and its excitations.

\textit{Acknowledgements}.\,---\,We thank P.~Gegenwart, A.~Tsirlin, H.~Rosner and C.~Geibel for fruitful discussions. Experiments at the ISIS Neutron and Muon Source were supported by a beamtime allocation from Science and Technology Facilities Council. We thank C.~Klausnitzer and H.~Rave for technical support during the magnetization and specific heat measurements at the MPI. Z. Z. and L. H. acknowledge the DFG for financial support (grant HO 4427/3). This work has been partially funded by the German Research Foundation (DFG) within the Collaborative Research Center SFB~1143 (projects B03, C02, C03, and A05).

\end{document}